\documentclass[11pt]{article}

\usepackage{amssymb,amsmath,latexsym, hyperref, cite, natbib, graphicx, refstyle, pgfplotstable, booktabs, rotating, color, tabularx, authblk, array, url, listings}

\usepackage{multirow}
\newcolumntype{C}[1]{>{\centering\arraybackslash}p{#1}}

\usepackage{geometry}
\usepackage{setspace}
\usepackage[multiple, hang,splitrule]{footmisc}


\usepackage[font=small, margin=0cm]{caption}

\usepackage[T1]{fontenc}
\begin{document}
\title{\textbf{Matching MEDLINE/PubMed Data with Web of Science (WoS): A Routine in \textit{R} language}}

\author[1]{\textbf{Daniele Rotolo}\thanks{Corresponding author: d.rotolo@sussex.ac.uk, Phone: +44 1273 872980}}
\author[2]{\textbf{Loet Leydesdorff}\thanks{loet@leydesdorff.net}}

\affil[1]{SPRU --- Science Policy Research Unit, University of Sussex}
\affil[2]{Amsterdam School of Communication Research (ASCoR), University of Amsterdam}
		
\date{\textit{Brief communication} --- Version: \today\linebreak 
	Accepted for publication in the\linebreak 
	\textit{Journal of the Association for Information Science and Technology}}

%-------------------------------------------------------------------------------
%	Abstract
%-------------------------------------------------------------------------------
\maketitle
\begin{abstract}
\onehalfspacing
\noindent We present a novel routine, namely \textit{medlineR}, based on \textit{R} language, that enables the user to match data from MEDLINE/PubMed with records indexed in the ISI Web of Science (WoS) database. The matching allows exploiting the rich and controlled vocabulary of Medical Subject Headings (MeSH) of MEDLINE/PubMed with additional fields of WoS. The integration provides data (e.g. citation data, list of cited reference, list of the addresses of authors' host organisations, WoS subject categories) to perform a variety of scientometric analyses. This brief communication describes \textit{medlineR}, the methodology on which it relies, and the steps the user should follow to perform the matching across the two databases. In order to specify the differences from \cite{Leydesdorff2013b}, we conclude the brief communication by testing the routine on the case of the "Burgada Syndrome".\newline\newline
\par
\end{abstract}
\clearpage

%-------------------------------------------------------------------------------
%	Introduction
%-------------------------------------------------------------------------------
\section{Introduction}

The use of Medical Subject Headings (MeSH) of the US National Library of Medicine (NLM) for scientometric analysis of the medical context has increased in the last few years. These include the use of MeSH terms to delineate domains \citep{Lundberg2006}, to identify emerging topics \citep{Ohniwa2010} and research areas \citep{Guo2011}, to map the dynamics of emerging technologies \citep{Leydesdorff2012}, to evaluate the impact of funding sources on the number of citations publications receive \citep{Boyack2004}, to perform co-word analyses \citep{Stegmann2003}, and to disambiguate author names \citep{Torvik2005}.

The MeSH classification, which is integrated in MEDLINE/PubMed, is a rich and controlled vocabulary generated through an intense indexing process performed by examiners. Terms, namely 'descriptors', are assigned to documents to delineate their content at different levels of specificity. The 2014 MeSH vocabulary is specifically composed by 27,149 descriptors which are organised in a tree-like structure.\footnote{The MeSH tree is organised in 16 branches representing different medical areas (e.g. diseases, chemical and drugs, therapeutic techniques). Additional details on the MeSH tree are available at \url{ww.nlm.nih.gov/pubs/factsheets/mesh.html}}  Descriptors may be also complemented with one or more 'qualifiers'. These terms further contextualise the meaning of the descriptors to which they are assigned in relation to the content of the considered document. 

The rich vocabulary provided by the MeSH classification can be used to delineate samples of documents in a number of medical areas and, as discussed, at different levels of specificity. For example, if one aims to examine the publication activity associated with a given disease, the MeSH descriptors associated with this disease can be  identified in the "MeSH browser interface"\footnote{\url{www.nlm.nih.gov/mesh/MBrowser.html}} and used to build a search string in the MEDLINE/PubMed interface. The retrieval of the set of associated documents is consequential. Using MeSH-based search strategy has been suggested as an approach preferable to using keywords in titles (and abstracts), journals, or authors' names for the delineation \citep{Lundberg2006}. 

Data from MEDLINE/PubMed however poses important limitations to scientometric analyses since they do not include key fields that are instead listed in commercial databases as ISI Web of Science (WoS) and SCOPUS. These include a publication's number of citations, list of cited references, as well as list of the addresses of authors' host organisations. For example, MEDLINE/PubMed only provides corresponding authors' addresses, while full information is required to investigate co-authorship networks at organisational level, to perform geographical mapping, and to aid the disambiguation of authors' names.\footnote{The NLM has recently announced a new policy to enable publishers to submit full information on authors' addresses starting from October 2013 (NLM Tech Bull. 2013 Sep-Oct;(394):b4).}

WoS and SCOPUS databases integrate the MeSH classification. WoS has a dedicated MEDLINE interface, while SCOPUS enables searching MeSH descriptors within the "Indexterms()" field. However, "Indexterms()" is not specific for the MeSH classification. These terms are assigned to SCOPUS records by indexers and integrated with thesauri that Elsevier owns of licenses. Searches within the "Indexterms()" field therefore extends also to records not included in MEDLINE/PubMed. WoS MEDLINE interface also has some limitations. It does not provide direct access to the above-listed key fields, i.e. the MEDLINE interface is 'weakly' integrated with the WoS core interface. The user has, for example, to enter record by record in WoS MEDLINE interface to collect citation data since they are not included in the download of the identified set of documents (based on WoS 5.13 version).

Following the lead of previous works \citep{Leydesdorff2013b}, this brief communication presents a novel routine, namely \textit{medlineR}, which enables the integration of data obtained by querying MEDLINE/PubMed with data available in WoS. The routine builds on some of the basic functions included in \textit{R}, a widely diffused open-source language and environment available across multiple platforms, and on the package "\textit{stringr}" \citep{Wickham2010}, which can be easily installed in the \textit{R} environment by the user. 

The \textit{medlineR} routine and methodological approach is described in the followings. We apply the routine to  the same case-study \cite{Leydesdorff2013b} investigated. This allows specifying differences between the two routines. We conclude the brief communication discussing the implications deriving from the integration of data across MEDLINE/PubMed and WoS.

%-------------------------------------------------------------------------------
%	Methods
%-------------------------------------------------------------------------------
\section{Methods}
To match data collected from MEDLINE/PubMed with WoS data by using \textit{medlineR}, the user has to follow a number of steps that are listed below. The code is reported in the Appendix and the \textit{medlineR} script is available at \url{http://www.danielerotolo.com/#!medliner/cid7}.

\begin{enumerate}

\item The user first installs \textit{R}\footnote{\textit{R} can be downloaded at \url{www.r-project.org}} and the package "\textit{stringr}" \citep{Wickham2010}.\footnote{Details on the package are available at \url{http://cran.r-project.org/web/packages/stringr/index.html}} The latter can be installed from the \textit{R} command line using the following string \textit{install.packages("stringr")}. An Internet connection is required.

\item The user can identify a list of "PubMed Identifiers", namely PMIDs, in MEDLINE/PubMed to match with WoS data. The list of PMID can be derived from the set of publications obtained from the specific query the user is performing. The interface of MEDLINE/PubMed allows for the download of PMIDs in \textit{txt} format.\footnote{\url{www.ncbi.nlm.nih.gov/pubmed}} It is worth noting that \textit{medlineR} can also work directly at WoS MEDLINE with a valid search string in the advanced-search interface. In this case, the user skips step 2 and step 3.

\item Assuming that the list is composed by $M$ PMIDs, the user builds an advanced search string in the WoS MEDLINE interface according to the following syntax: $PM=PMID_1\; or \;PM =PMID_2\;or\;...\;PM=PMID_M$. Conventional spreadsheets can be used to generate the search string. 

\item This search string is then used to query WoS MEDLINE through the advanced search interface. The query will return a list of documents. Each document can be accessed through a weblink. This implies that a \textit{url} link is associated to each document the search retrieved. 

\item The user has to input three parameters in the \textit{medlineR} script: 
\begin{enumerate}

\item One of the \textit{url} links to WoS MEDLINE documents (variable \textit{wosurl}). To do so, the user can access to the first document in the list and copy-paste, between quotation marks, the \textit{url} link associated with this document in the \textit{R} code --- \textit{medlineR} will use this \textit{url} to generate the remaining ones automatically.\footnote{If the user builds a search string in the advanced search interface of WoS MEDLINE and the search returns more than 10 documents, the \textit{url} associated to one of these documents will have the following format: "\url{http://apps.webofknowledge.com/full_record.do?product=MEDLINE&search_mode=AdvancedSearch&qid=***&SID=***&page=***&doc=***}". Stars replace codes that are generated by WoS MEDLINE according to the performed search and the session number. In the case of a search in the general interface the \textit{url} will have the following format: "\url{http://apps.webofknowledge.com/full_record.do?product=MEDLINE&search_mode=GeneralSearch&qid=***&SID=***&page=***&doc=***}".}

\item The number of documents to collect (variable \textit{numdocs}).

\item The path to the folder in which the outputs of the \textit{medlineR} routine should be saved. This should be inputted in \textit{R} between quotation marks (e.g. "C:\textbackslash\textbackslash Users \textbackslash\textbackslash user" in Windows or "/Users/username/Desktop/" in Mac OS X).\footnote{\textit{R} requires double backslash when the path to the selected folder is specified in Windows.}
\end{enumerate}

\item The user has to launch \textit{medlineR} (the shortcut in Windows is "CTRL+A" and then "CTRL+R" whereas in Mac OS X is "CMD+A" and then "CMD+Return"), which parses the \textit{wosurl} variable to generate the whole set of links pointing to the identified documents and collect the full \textit{html} code of the associated webpages. The \textit{html} code is then parsed to retrieve documents' UTs, i.e. "Unique Article Identifiers" in WoS. An indication of the number of processed records is reported in the \textit{R} interface as the routine advances.

\item The \textit{medlineR} routine generates three different outputs: (i) a set of files (sequentially named \textit{wosPMID=1.txt, wosPMID=2.txt, etc.})  including the \textit{html} code of each webpage, (ii) a document, called \textit{wosut.txt}, which lists PMIDs with associated (when available) UTs, and (iii) a file, called \textit{search.txt}, that provides the full search string for WoS as generated from the collected UTs.\footnote{The \textit{medlineR} routine may return warning messages at the end of the data collection process. The user can ignored these messages since they do not affect the produced files.} This string can be use in the advanced interface of WoS to retrieve the full records of the identified documents. These can be then downloaded from the WoS interface.

\end{enumerate}

%-------------------------------------------------------------------------------
%	A case-study
%-------------------------------------------------------------------------------
\section{A case-study: The Brugada Syndrome}
For a comparative analysis with the results obtained by \cite{Leydesdorff2013b}, we applied the \textit{medlineR} routine on the case-study of the Brugada Syndrome (BRs) that is a rare cardiac disease --- for more details on the case-study see \cite{Leydesdorff2013b}. BRs is identified in the MeSH classification with the term "Brugada Syndrome", which is coded in the tree with "C14.280.067.322" as a cardiac arrhythmia and "C16.320.100" as a congenital disease. This term has been indexed since 2007.

We searched for all publications to which the "Brugada Syndrome" MeSH term was assigned during the 2010-2011 period. We performed the search in MEDLINE/PubMed on 6 June 2014 and compared the results obtained from WoS MEDLINE interface and SCOPUS by using the same search approach. \Tabref{databases} summarises the results. MEDLINE/PubMed returned 349 records while WoS MEDLINE and SCOPUS returned lower and higher number of records, respectively. This shows evidence of the limitations of applying MeSH-based searches directly on those databases.

We launched \textit{medlineR} to match the records obtained from the search performed in  MEDLINE/PubMed with those that are listed in WoS MEDLINE and then in the core interface of WoS. The routine identified 294 UTs out the 349 records included in the sample. As a comparison, we also used the \textit{medline.exe} routine developed by \cite{Leydesdorff2013b} the same day we used \textit{medlineR}. The number of retrieved UTs was identical. 

The \textit{medlineR} routine does not requires large computational power or memory since most of the collected data are written on the hard drive as they are collected. The main parameters that affect its performance are the number of records to collect and the quality of the Internet connection. We used the routine on a Macbook Pro with a 2.8 GHz Intel Core i7 processor and 8GB (1333MHz DDR3) RAM. The data collection for the BRs case-study was achieved in about 11 minutes.

%=====
\setlength{\tabcolsep}{10pt}
\renewcommand{\arraystretch}{1.5}
\begin{table}\footnotesize
	\caption{\label{tab:databases}Comparing databases: Records related to the "Brugada Syndrome" (2010-2011 period).}
	\centering
{\begin{tabular}{lp{9.5cm}c}
\hline\hline
\textbf{Database} &  \textbf{Search string} &  \textbf{Records}\\
\hline
MEDLINE/PubMed & "Brugada syndrome"[MeSH Terms] AND ("2010.01.01"[PDAT]: "2011.31.12"[PDAT]) & 349\\
MEDLINE WoS & MH="Brugada Syndrome" AND (PY=2010 OR PY=2011) & 323\\
SCOPUS & INDEXTERMS("Brugada Syndrome") AND ((PUBYEAR = 2010) OR (PUBYEAR = 2011)) & 591 \\
\hline\hline
\multicolumn{3}{l}{Note: The searches were performed on 6th June 2014.} \\
\end{tabular}
}
\end{table}
%=====

%-------------------------------------------------------------------------------
%	Conlcusion
%-------------------------------------------------------------------------------
\section{Conclusions}
The rich and controlled vocabulary of the MeSH classification of MEDLINE/PubMed allows for rapid delineation of publication samples in medical areas at different levels of specificity. More than 27,000 terms (descriptors) populate the classification and the list of those is also constantly updated with new terms to cover emerging areas. MeSH therefore provides a valuable alternative to searches of keywords in the titles and abstracts of publications or to searches that relies on sets of journal titles. However, MEDLINE/PubMed data poses limitations to the scientometric analysis. Key fields required for scientometrics are missed or incomplete. For example, MEDLINE/PubMed does not list the references a publication cited neither future citations, which in turn are key inputs for bibliographic coupling and citations analyses, respectively. In addition, authors' addresses are reported only for the corresponding author, thus limiting the possibility to examine inter-organisational networks, to perform geographical mapping, as well as to support the disambiguation of author names. The missing data also does not allow to perform scientometric mapping with a number of available mapping tools (e.g. the map of science based on WoS subject categories, the journal mapping) \citep[for an overview, see][]{Rotolo2014}.

These fields are available in commercial databases as WoS and SCOPUS, which also include interfaces to build MeSH-based searches to retrieve publication data. Yet, in the case of WoS,  searches are not comprehensive whereas, in the case of SCOPUS, searches extend also to documents not included in MEDLINE/PubMed. In this brief communication, we proposed a novel routine in \textit{R} language, namely \textit{medlineR}, that enables users to integrate, by using the WoS MEDLINE interface, data from MEDLINE/PubMed with the above mentioned fields available in WOS. \textit{MedlineR} specifically matches PMIDs from MEDLINE/PubMed with UTs from WoS. The \textit{R} language on which \textit{medlineR} is based also enables the user to introduce additional functionalities such the parsing and retrieval of other fields available in the original \textit{html} code of WoS or rapid analyses of the collected data directly in the \textit{R} environment. The code is indeed editable and adaptable to the specific requirements of the scientometric analysis the user aims to undertake.

%-------------------------------------------------------------------------------
%	Acknowledgements
%-------------------------------------------------------------------------------
\section*{Acknowledgements}
We acknowledge the support of the UK Economic and Social Research Council (award RES-360-25-0076 -  \href{http://www.interdisciplinaryscience.net/mdetp}{"{\color{blue}Mapping the Dynamics of Emergent Technologies}"}). Daniele Rotolo also acknowledges the support of the People Programme (Marie Curie Actions) of the European Union's Seventh Framework Programme (FP7/2007-2013) (award PIOF-GA-2012-331107 - \href{http://www.danielerotolo.com/#!netgenesis/c8kr}{"{\color{blue}NET-GENESIS: Network Micro-Dynamics in Emerging Technologies}}"). We are grateful to Carlos Benito, Francois Perruchas, and Ismael Rafols for their feedbacks.
%-------------------------------------------------------------------------------
%	References
%-------------------------------------------------------------------------------
\bibliographystyle{apalike}

 \clearpage
%-------------------------------------------------------------------------------
%	Appendix
%-------------------------------------------------------------------------------
\section*{Appendix}
\begin{singlespace} 
\begin{lstlisting}[caption={\textit{medlineR} script}, label=code]

# medlineR (v.1.0): Matching MEDLINE/PubMed and ISI Web of Science (WoS)
# Rotolo and Leydesdorff (2014)
# Please refer to the fair use policy at http://wos.isitrial.com/policy/Policy.htm

# SETTING PARAMETERS - - - - - - - - - - - - - - - - - - - - - - - - - - - - - - - 

# insert the WoS (MEDLINE interface) link including quotes
wosurl<-"..."

# insert the number of document to match
numdocs<-...

# setting the output folder including quotes
# (e.g. "C:\\Users\\user" in Windows or 
# "/Users/username/Desktop/" in Mac OS X)
setwd("...")

#- - - - - - - - - - - - - - - - - - - - - - - - - - - - - - - - - - - - - - - - -   

# loading library
library(stringr)                  

# dowloading the html code and parsing data
wosurl_str<-substr(wosurl, 1, (nchar(wosurl)-1))

for(k in 1:numdocs)
{
  print(paste("--Record number: ",k," (out ",numdocs,")--",sep=''))
  url<-paste(wosurl_str,k,sep='')
  all_lines<-readLines(url)
  line<-all_lines[str_detect(all_lines,"UT=WOS:")]
  line_str<-unlist(strsplit(str_extract(line,"UT=WOS:.+"),"&"))
  wosut<-sub('WOS:','',line_str[1])
  
  line<-all_lines[str_detect(all_lines,"NCBI_DB&PMID")]
  line_str<-unlist(strsplit(str_extract(line,"PMID.+"),"&"))
  pmid<-line_str[1]
  
  data<-cbind(pmid,wosut)
  
  write.table(data,file='wosut.txt',row.names=F,col.names=F,append=T,sep=",")
  write.table(all_lines,file=paste('wos',pmid,'.txt',sep=''),row.names=F,col.names=F,append=F)
}

# creating the search string for WoS
uts<-read.csv(file="wosut.txt",header=F,sep=",",fill=T)
uts<-subset(uts,uts[,2]!="")
searchwos<-paste(uts[,2],collapse=' OR ')
write.table(searchwos,file='search.txt',row.names=F,col.names=F,append=F,quote=F)  
#- - - - - - - - - - - - - - - - - - - - - - - - - - - - - - - - - - - - - - - - - - - 
\end{lstlisting}
\end{singlespace} 

\end{document}